\documentclass[aps,prl,reprint]{revtex4-2}
\usepackage{amsmath,amssymb}
\usepackage{hyperref}
\usepackage{graphicx}% Include figure files
\usepackage{dcolumn}% Align table columns on decimal point
\usepackage{bm}% bold math
\usepackage{hyperref}

\begin{document}
\preprint{APS/123-QED}
\title{Stacking theory for bilayer two-dimensional magnets}

\author{Jun-Xi Du}
\thanks{These authors contributed equally to this work.}
\author{Sike Zeng}
\thanks{These authors contributed equally to this work.}

\author{Yu-Jun Zhao}
\email{zhaoyj@scut.edu.cn}
\affiliation{Department of Physics, South China University of Technology, Guangzhou 510640, China}

\date{\today}

\begin{abstract}
Two-dimensional unconventional magnetism has recently attracted growing interest due to its intriguing physical properties and promising applications in spintronics. However, existing studies on stacking-induced unconventional magnetism mainly focus on specific materials and stacking configurations. Here, we develop a general symmetry-based stacking theory for two-dimensional magnets. We first introduce spin layer groups as the fundamental symmetry framework, providing the essential magnetic symmetry information for the stacking theory. Based on this framework, we construct the complete set of 448 collinear spin layer groups for describing two-dimensional collinear magnets. Subsequently, we develop a general magnetic stacking theory applicable to \textit{arbitrary} magnetic systems and derive its general solutions. Using CrF$_3$ as an illustrative example, we show how this theory enables designs of two-dimensional unconventional magnetism, as validated by first-principles calculations. We realize two-dimensional fully compensated ferrimagnetism through our stacking theory. Our work provides a general symmetry-guided platform for discovering and designing stacking-induced unconventional magnetism.
\end{abstract}

\maketitle
\textit{Introduction---}Magnetism, as one of the most fundamental and dynamic fields in condensed matter physics, has long attracted intense research interest owing to its rich underlying physics and its pivotal role in modern applications. Recently, unconventional magnetic materials that integrate the advantages of both ferromagnetism and antiferromagnetism have attracted considerable attention, as they hold strong promise as a foundation for next-generation spintronic devices\cite{liu2025different}. Meanwhile, the experimental realization of monolayer CrI$_3$\cite{huang2017layer} and bilayer Cr$_2$Ge$_2$Te$_6$\cite{gong2017discovery}, together with the successful observation of long-range magnetic order therein, has stimulated increasing interest in two-dimensional (2D) magnetism\cite{wangMagneticGenomeTwoDimensional2022}. From the perspective of applications, 2D magnetic materials, in contrast to bulk systems, possess high tunability and can be readily integrated into van der Waals (vdW) heterostructures, thereby offering versatile platforms for next-generation spintronic devices\cite{mak2019probing}. In this context, the extension of unconventional magnetism into the 2D limit has become a rapidly growing research direction. Several classes of two-dimensional unconventional magnetism, including 2D altermagnets\cite{PhysRevX.12.031042,PhysRevB.110.054406}, 2D fully compensated ferrimagnets\cite{PhysRevLett.134.116703}, and type-IV 2D collinear magnets\cite{rn1l-d6cq}, have been proposed recently and have attracted considerable interest. These systems exhibit time-reversal symmetry-breaking responses even in the absence of a macroscopic magnetization. From a symmetry perspective, such magnetic states can be  identified using spin-group symmetries. So far, research on 2D unconventional magnetism is still at an early stage, and corresponding experimental realizations remain largely unexplored\cite{zengClassificationDesignTwodimensional2026}.

Compared with bulk materials, 2D systems offer a unique `stacking' degree of freedom, enabling efficient manipulation of crystalline symmetries\cite{PhysRevLett.130.146801}. Consequently, stacking bilayer or multilayer structures has become a powerful route to realize 2D unconventional magnetism\cite{heNonrelativisticSpinMomentumCoupling2023,sheoranNonrelativisticSpinSplittings2024,zengBilayerStacking$A$type2024,panGeneralStackingTheory2024,PhysRevLett.133.206702, PhysRevB.110.224418, guoSpinOrderingInduced2025}, with stacking-induced 2D altermagnetism receiving the most intensive attention. Nevertheless, previous investigations remain largely restricted to specific materials, and the theory developed for A-type altermagnets\cite{zengBilayerStacking$A$type2024,panGeneralStackingTheory2024} fails to capture other altermagnets as well as other categories of 2D unconventional magnetism. These limitations naturally raise a fundamental question: is there a general theory of magnetic stacking to guide the realization of 2D unconventional magnetism? Such a theory must resolve two central problems: first, given a monolayer and a specific stacking operation, predicting the resulting magnetic properties of the bilayer; second, given a monolayer and a target magnetic state, identifying the requisite stacking operations. This general theory becomes possible owing to recent developments in spin-group theory\cite{liuSpinGroupSymmetryMagnetic2022a,PhysRevX.14.031037,PhysRevX.14.031038,PhysRevX.14.031039}. In contrast to magnetic groups, spin groups capture finer details of magnetic geometry, offering a powerful foundation for the theoretical design of magnetic materials.

\begin{figure*}
	
	\centering
	\includegraphics[width=\linewidth]{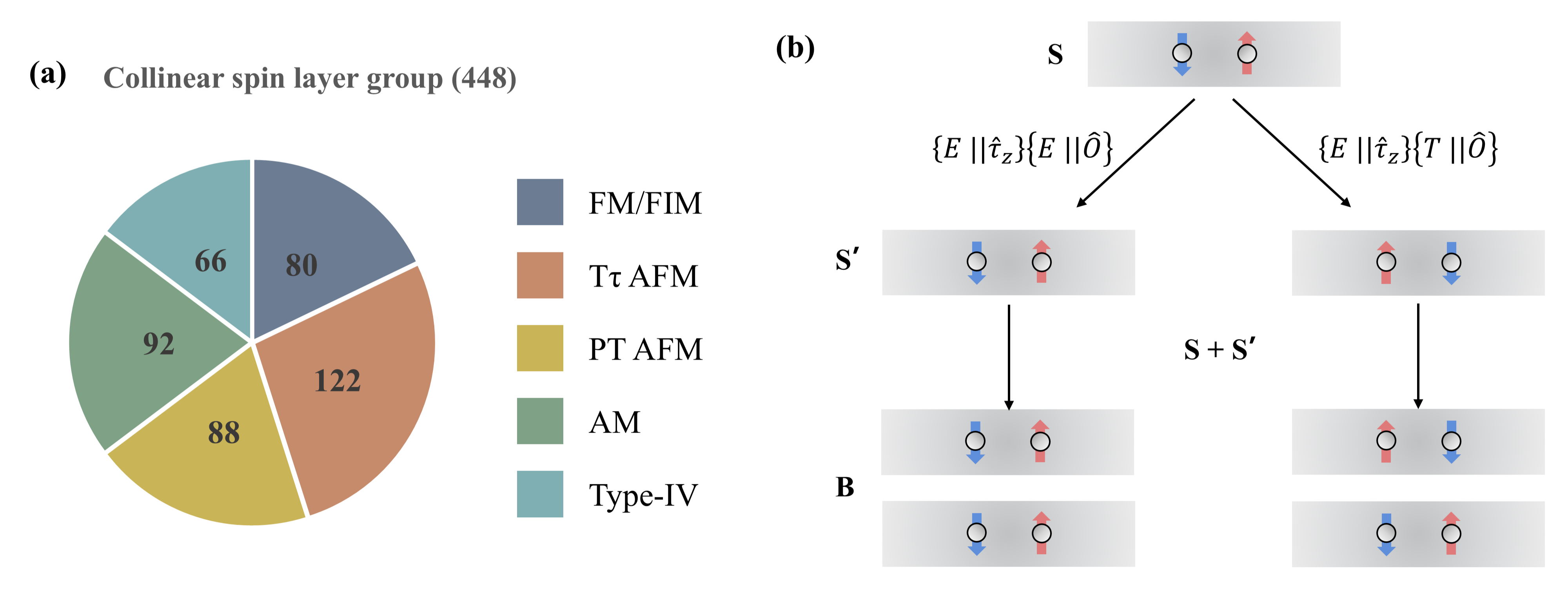}
	\caption{\label{fig:figure1}Classification of collinear spin layer groups and schematic illustration of the bilayer stacking in magnetic materials. (a) The 448 collinear spin layer groups (cSLGs) are classified into five categories according to the magnetic states they describe, including 80 ferromagnetic or ferrimagnetic (FM/FIM), 122 $\mathcal{T}\boldsymbol{\tau}$ antiferromagnetic ($\mathcal{T}\boldsymbol{\tau}$ AFM), 88 $\mathcal{P}\mathcal{T}$ antiferromagnetic ($\mathcal{P}\mathcal{T}$ AFM), 92 altermagnetic (AM) and 66 type-IV 2D collinear magnetic (Type-IV) cSLGs. (b) Starting from a monolayer $S$, a stacking operation $\{E \,\|\, \hat{\tau}_z\} \{c \,\|\, \hat{O}\}$ is applied to generate $S'$, which is then stacked onto $S$ to form a bilayer $B$.}
	
\end{figure*}

In this letter, we establish a general stacking theory for bilayer two-dimensional magnets by integrating spin-group symmetry with the freedom degree of stacking. We begin by introducing the spin layer group and constructing the complete set of 448 collinear spin layer groups. Subsequently, we develop the stacking theory for 2D magnets together with its analytical solutions. Using CrF$_3$ as a prototype, we demonstrate how this theory enables predictive design of unconventional magnetic states, with first-principles calculations confirming the theoretical predictions. Moreover, through symmetry analysis, we reveal that monolayer CrF$_3$ is a type-IV two-dimensional collinear magnet. This symmetry-based framework provides unified guidelines for engineering magnetism in 2D systems, opening new avenues for stacking-induced unconventional magnetism.

\textit{Spin layer groups---}As the foundation of the magnetic stacking theory, spin layer groups (SLGs) are introduced first. A symmetry operation of a SLG can be expressed as $\{ R_i \,\|\, \hat{R}_j \}$, where $R_i$ acts in spin space and $\hat{R}_j$ is a layer-group operation in real space. In this letter, a caret is used to distinguish space-group symmetries from point-group symmetries. A spin layer group can be decomposed as $G = G_{\mathrm{SO}} \times G_{\mathrm{NSL}}$, where $G_{\mathrm{SO}}$ is the spin-only group containing only pure spin operations $\{ R_i \,\|\, E \}$, and $G_{\mathrm{NSL}}$ denotes the nontrivial spin layer group\cite{litvinSpinPointGroups1977a}. For collinear spin layer groups (cSLGs) describing 2D collinear magnets, the spin-only group is the internal semidirect product of $\mathrm{SO}(2)$ and $\mathrm{Z}_2^K$, $G_{\mathrm{SO}} = \mathrm{SO}(2) \rtimes \mathrm{Z}_2^K$, where $\mathrm{SO}(2)$ describes continuous spin rotations about the collinear axis and $\mathrm{Z}_2^K \equiv \{ E, U_{\textbf{n}}(\pi)\mathcal{T} \}$\cite{liuSpinGroupSymmetryMagnetic2022a}. Here, $U_{\textbf{n}}(\pi)$ denotes a $180^\circ$ spin rotation about an axis perpendicular to the spin, and $\mathcal{T}$ is the time-reversal operator. For coplanar SLGs, the spin-only group is given by $G_{\mathrm{SO}} = \{ E, U_{\textbf{n}}(\pi)\mathcal{T} \}$, where $\textbf{n}$ denotes the direction perpendicular to the plane of the magnetic structure. For noncoplanar SLGs, the spin-only group is only the identity, i.e. $G_{\mathrm{SO}} = \{E\}$.

We then provide the complete set of 448 cSLGs that characterize 2D collinear magnets. The nontrivial part of a cSLG is in one-to-one correspondence with a magnetic layer group (MLG), owing to their similarity in mathematical structure\cite{chenUnconventionalMagnonsCollinear2025}, as listed in Table S1\cite{NOTE}. Since a collinear spin arrangement allows only two opposite spin orientations, the spin-space part of the nontrivial SLG is restricted to the identity $E$ and time reversal $\mathcal{T}$. For collinear ferromagnets, which possess a single magnetic sublattice, the nontrivial cSLG takes the form $\{ E \,\|\, G \}$, corresponding to type-I MLGs. For collinear antiferromagnets, which possess two magnetic sublattice, the nontrivial cSLG takes the form $\{ E \,\|\, H \}+\{ \mathcal{T} \,\|\, AH \}$, corresponding to type-III and type-IV MLGs. Here, $H$ denotes an index-2 normal subgroup of $G$, and $A$ is a coset representative. Based on this correspondence with MLGs\cite{litvinMagneticGroupTables2013b}, we enumerate all nontrivial parts of cSLGs, totaling 448, as listed in Tables S3--S7\cite{NOTE}. 

We further classify these cSLGs according to the magnetism they describe, as illustrated in Fig. \ref{fig:figure1}(a). First, the 80 cSLGs of the form $\{ E \,\|\, G \}$ describe ferromagnets, as well as ferrimagnets in which multiple magnetic sublattices exist but are not related by symmetry. 2D fully compensated ferrimagnets are also described by these 80 cSLGs. As discussed above, the nontrivial spin layer group describing collinear antiferromagnets takes the form $\{ E \,\|\, H \} + \{ \mathcal{T} \,\|\, A H \}$. When $A$ is a pure fractional translation $\boldsymbol{\tau}$, these cSLGs describe $\mathcal{T}\boldsymbol{\tau}$ antiferromagnets, totaling 122. There are 88 cSLGs possess $\mathcal{P}\mathcal{T}$ symmetry, where $\mathcal{P}$ denotes spatial inversion, but lack $\mathcal{T}\boldsymbol{\tau}$ symmetry.  We classify these as $\mathcal{P}\mathcal{T}$ antiferromagnets. When $A$ is neither in $\{\boldsymbol{\tau}, \mathcal{P}\}$ but belongs to $\{ C_{2z}, M_z \}$, the corresponding cSLGs describe type-IV 2D collinear magnets, totaling 66. Finally, when $A$ is not in $\{\boldsymbol{\tau}, \mathcal{P}, C_{2z}, M_z\}$, these cSLGs correspond to 2D altermagnets, totaling 92. Altermagnets exhibit spin-splitting bands in the nonrelativistic limit, whereas the other antiferromagnets has spin-degenerate bands.

\textit{Magnetic stacking theory---}Subsequently, we present the magnetic stacking theory, which integrates spin layer groups with the freedom of stacking for 2D materials\cite{PhysRevLett.130.146801}. Since the interlayer vdW interactions are typically weak, we assume that stacking does not alter the intralayer magnetic order, which remains the same as in the corresponding monolayer. As illustrated in Fig. \ref{fig:figure1}(b), a magnetic monolayer $S$ with spin layer group $G_S = \{\{s \,\|\, \hat{R}_S\}\}$ generates a second layer $S'$ through the stacking operation $\{E \,\|\, \hat{\tau}_z\} \{c \,\|\, \hat{O}\}$. Here, $\{s \,\|\, \hat{R}_S\}$ represents a symmetry operation of the monolayer, with $s$ and $\hat{R}_S$ acting in spin space and real space, respectively. The operation $\{E \,\|\, \hat{\tau}_z\}$ is a pure translation along $z$ axis and does not affect the symmetry of bilayer. In contrast, $\{c \,\|\, \hat{O}\} \equiv \{c \,\|\, O \mid \tau_O\}$ represents the nontrivial part of the stacking operation, where $c$ denotes the spin-space part, and $O$ and $\tau_O$ are the point-group operation and the translation in real space, respectively. The resulting bilayer $B$, characterized by the spin layer group $G_B  = \{\{b \,\|\, \hat{R}_B\}\}$, is composed of these two layers. Here, $b$ denotes the spin-space part of the spin-group symmetry of the bilayer. Based on the above analysis, we can formulate a magnetic bilayer stacking model
\begin{equation}
	\left\{\label{eq:1}
	\begin{aligned}
		S &=\, ^s\!\hat{R}_S S, \ \ \forall\, ^s\!\hat{R}_S \in G_S \\
		B &=\, ^b\!\hat{R}_B B, \ \ \forall\, ^b\!\hat{R}_B \in G_B\\
		B &= S+S' \\
		S'&=\, ^1\hat{\tau}_z ^c\!\hat{O} S
	\end{aligned}
	\right.
\end{equation}
where we denote the spin-group operations in a more compact form for simplicity, such as $^s \! \hat{R}_S \equiv \{s \,\|\, \hat{R}_S\}$. Note that the SLGs $G_S$ and $G_B$ include not only the nontrivial part but also spin-only group elements.

We now turn to the first problem: given a monolayer and a specified stacking operation, how can one predict the SLG of the resulting bilayer system? Depending on whether the $z$ coordinate is reversed, the spin-group symmetries of the bilayer can be divided into two classes, $^b \!\hat{R}_B^{+}$ and $^b \!\hat{R}_B^{-}$, which preserve and reverse the $z$ coordinate, respectively. From a physical perspective, $^b \!\hat{R}_B^{-}$ interchanges $S$ and $S'$, while $^b \!\hat{R}_B^{+}$ does not. For example, the spin-group operation $\{E \,\|\, P \mid \mathbf{0}\}$ interchanges the two monolayers and is thus classified as  $^b \!\hat{R}_B^{-}$. The SLG of a bilayer is composed of two sets, $\{{}^{b}\!\hat{R}_B^{+}\}$ and $\{{}^{b}\!\hat{R}_B^{-}\}$, i.e. $G_B= \{{}^{b}\!\hat{R}_B^{+}\} \cup \{{}^{b}\!\hat{R}_B^{-}\}$. Using Eqs. (\ref{eq:1}), we can  obtain
\begin{equation}
	\label{eq:2}
	\{{}^{b}\!\hat{R}_B^{+}\} = G_S \cap\, ^c\hat{O}G_S\,^{c^{-1}}\!\!{\hat{O}}^{-1}
\end{equation}
and
\begin{equation}
	\label{eq:3}
	\{{}^{b}\!\hat{R}_B^{-}\} =\,^c\hat{O}G_S \cap G_S\,^{c^{-1}}\!\!{\hat{O}}^{-1}
\end{equation}
where ${}^{c^{-1}}\!\!\hat{O}^{-1}$ denotes the inverse of ${}^{c}\hat{O}$. The detailed derivation is presented in Sec. II of the Supplemental Material\cite{NOTE}. The spin-only group of $G_S$ and the stacking operation determine the spin-only group of $G_B$, which implies whether the bilayer magnetic configuration is collinear, coplanar, or noncoplanar. Therefore, if the class of the magnetic configuration in bilayer is known, the spin-only group of $G_S$ can be neglected. Eqs. (\ref{eq:2}) and (\ref{eq:3}) indicate that the SLG of bilayer is determined not only by the stacking configuration ($\hat{O}$), but also by the interlayer magnetic order ($c$).

\begin{figure*}
	
	\centering
	\includegraphics[width=\linewidth]{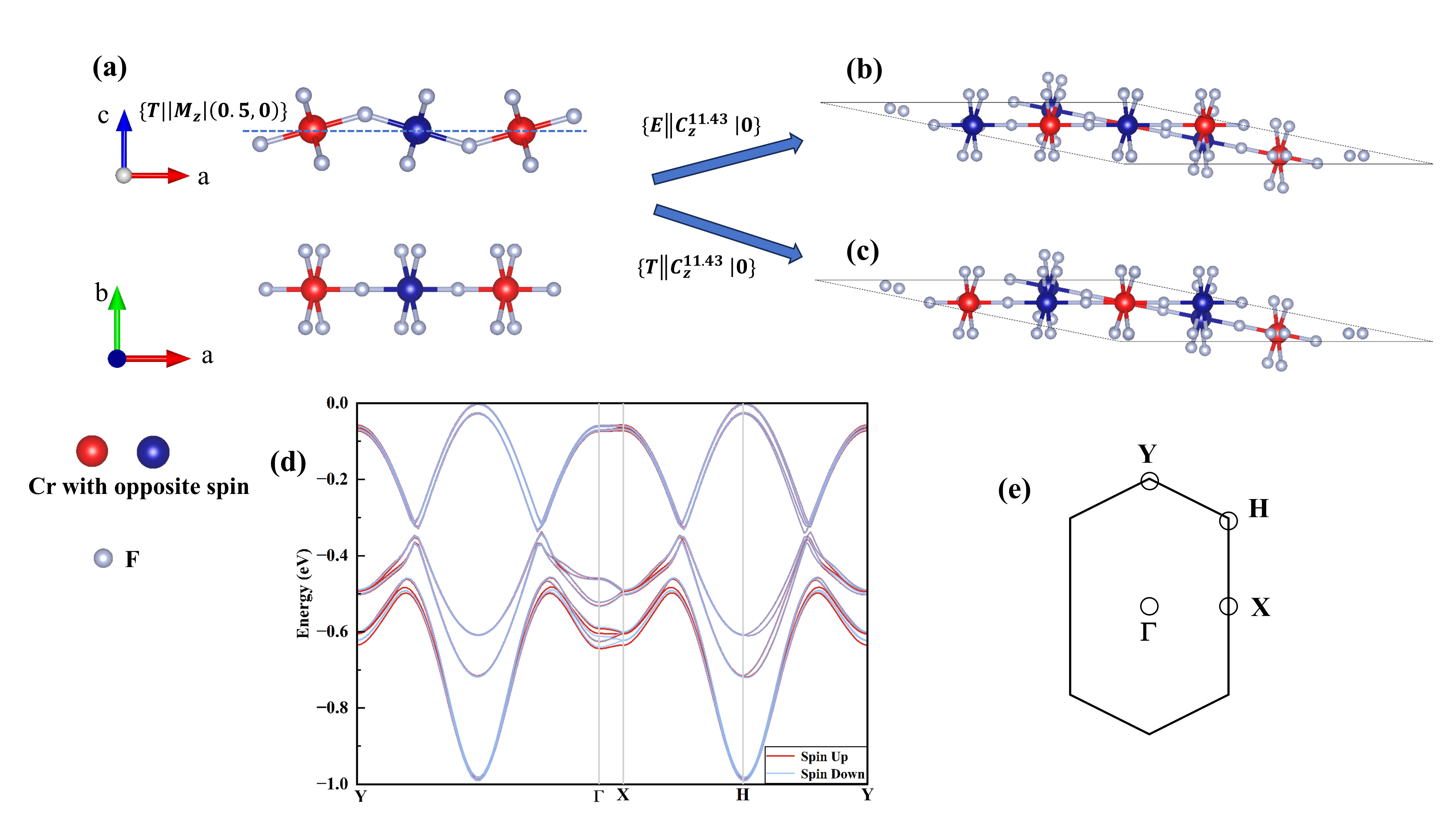}
	\caption{\label{fig:figure2}The crystal structures of monolayer CrF$_3$ and twisted bilayer CrF$_3$ with a twist angle of $11.43^\circ$, together with the band structure of bilayer CrF$_3$. (a) The crystal structures of monolayer CrF$_3$. A monolayer CrF$_3$ is stacked into a bilayer via stacking operations$\{c\,\|\, C^{11.43}_z \mid \mathbf{0}\}$, where $C^{11.43}_z$ denotes a rotation of $11.43^\circ$ about the $z$ axis. For $c = E$, the resulting bilayer CrF$_3$ (b) hosts fully compensated ferrimagnetism, whereas for $c = T$, the bilayer CrF$_3$ (c) exhibits altermagnetism. (d) Band structure of the twisted bilayer CrF$_3$ shown in (c), with the high-symmetry points in reciprocal space shown in (e).}
	
\end{figure*}

We then turn to the second problem: determining the stacking operations that yield a target bilayer magnetic state for a given monolayer. In the following, We still discuss ${}^{b}\!\hat{R}_B^{-}$ and ${}^{b}\!\hat{R}_B^{+}$ separately and the detailed derivations are provided in Sec. II of the Supplemental Material\cite{NOTE}. The spin-group symmetry ${}^{b}\!\hat{R}_B^{-}$ is allowed in the bilayer only if $({}^{b}\!\hat{R}_B^{-})^2 \in G_{S}$. This condition is a prerequisite for a given monolayer to be stacked into a bilayer with a specific symmetry ${}^{b}\!\hat{R}_B^{-}$. Once the above condition is satisfied, ${}^{b}\!\hat{R}_B^{-}$ can be generated or preserved in $B$ through a stacking operation ${}^{c}\hat{O} \in {}^{b}\!\hat{R}_B^{-} G_S$. In other words, selecting a stacking operation ${}^{c}\hat{O} \notin {}^{b}\!\hat{R}_B^{-} G_S$ can lead to the breaking of ${}^{b}\!\hat{R}_B^{-}$ in $B$. 

For ${}^{b}\!\hat{R}_B^{+}$, a necessary condition for its presence in the bilayer is that the monolayer already possesses the same symmetry, i.e. ${}^{b}\!\hat{R}_B^{+} \in G_S$. Consequently, stacking cannot generate ${}^{b}\!\hat{R}_B^{+}$; it can only preserve or break this symmetry, consistent with Eq. (\ref{eq:2}). Different ${}^{b}\!\hat{R}_B^{+}$ symmetries exhibit different dependencies on the stacking operation. For ${}^{b}\hat{C}_{nz}$, where $C_{nz}$ denotes a rotation about the $z$ axis, its preservation in $B$ is determined by the spin part ($c$) and the translational part ($\tau_O$), but not by the rotational part ($O$). By contrast, ${}^{b}\!\hat{E}$ (e.g., $\mathcal{T}\tau$) is determined by $c$ and $O$, but not by $\tau_O$. The symmetry is broken when the translational direction of ${}^{b}\!\hat{E}$ in the second layer does not align with the direction of any spin spiral in the first layer. For ${}^{b}\!\hat{M}_{\beta}$, where ${M}_{\beta}$ represents a mirror plane perpendicular to the material plane, all three components, $c$, $O$, and $\tau_O$, play a role. All operations in the spin-only group belong to ${}^{b}\!\hat{R}_B^{+}$ and thus can only be preserved or broken. Their preservation is determined solely by the spin-space part $c$ of the stacking operation. Crucially, this operation also determines whether the magnetic configuration in the bilayer is collinear, coplanar, or noncoplanar. These results demonstrate that stacking offers versatile means to manipulate symmetry, including interlayer magnetic order ($c$), sliding ($\tau_O$), and relative rotational or mirror alignments between monolayers ($O$).

We next consider several special cases. For the collinear case, where both the monolayer and the stacked bilayer exhibit collinear magnetic configurations, the spin part $c$ is restricted to $\{E, \mathcal{T}\}$. In this case, the realization of ${}^{b}\!\hat{R}_B^{+}$ becomes independent of $c$. Consequently, ${}^{b}\hat{C}_{nz}$ can only be broken by sliding, whereas ${}^{b}\!\hat{E}$ can only be broken by proper or improper rotations. In other cases, tuning the interlayer magnetic order provides an effective means to modulate ${}^{b}\!\hat{R}_B^{+}$. When only the breaking of  ${}^{b}\! R_B^{-}$ is of interest, such as the breaking of $\mathcal{PT}$ symmetry, a simple strategy is to choose ${}^{c}O \notin {}^{b}\! R_B^{-} G_{S0}$. Here, ${}^{b}\!R_B^{-}$ (${}^{c}O$) denotes the spin-point-group part of ${}^{b}\!\hat{R}_B^{-}$ ($^c\hat{O}$), and $G_{S0}$ is the spin point group associated with the $G_S$.

\textit{Unconventional magnetism in CrF$_3$---}Recently, an antiferromagnet CrF$_3$ with a rectangular crystal structure has been proposed\cite{chenDiscoveryUltrastableAntiferromagnetic2025}, as illustrated in Fig. \ref{fig:figure2}(a), which is found to be energetically more stable than the widely assumed hexagonal phase. Through symmetry analysis, we find that monolayer CrF$_3$ is a Type-IV 2D collinear magnet with the SLG $P\,^{-1}m\, ^1m\, ^{-1}a $, whose symmetry operations are listed in Table S6\cite{NOTE}. With the easy axis along the out-of-plane direction\cite{chenDiscoveryUltrastableAntiferromagnetic2025}, the system belongs to the magnetic group $Pmm'a'$ that permits a finite net magnetization along the $a$ direction. First-principles calculations including spin-orbit coupling (SOC) further reveal a small canting of the magnetic moments toward the $a$ direction. Due to the two-dimensional nature of CrF$_3$, the only experimentally relevant component of the anomalous Hall conductivity is $\sigma_{xy}$. However, based on magnetic group analysis, $\sigma_{xy}$ is forbidden by $\mathcal{M}_z\mathcal{T}$ in this system. Consequently, monolayer CrF$_3$ does not exhibit an observable anomalous Hall effect. 

We then investigate stacking bilayer CrF$_3$ as a platform for realizing other types of 2D unconventional magnetism. First, we discuss altermagnetism, which has recently attracted considerable attention. To realize altermagnetism, the symmetries $\{ \mathcal{T} \| \mathcal{A} \}$ must be broken, where $\mathcal{A} \in \{ \mathcal{P}, \mathcal{M}_z, \mathcal{C}_{2z}, \boldsymbol{\tau} \}$. Since the translational components associated with $\mathcal{P}$ and $\mathcal{M}_z$ are irrelevant in the present context, this requirement can be satisfied by choosing ${}^{c}O$ such that it belongs neither to $\{\mathcal{T}\|\mathcal{P}\}G_{S0}$ nor to $\{\mathcal{T}\|\mathcal{M}_z\}G_{S0}$, thereby ensuring the absence of these two symmetry operations in the bilayer. However, because the point group of rectangular lattice is $mmm$ and $c \in \{E, \mathcal{T}\}$, stacking operations that preserve the same magnetic unit-cell size for both the bilayer and the monolayer cannot realize altermagnetism. Because monolayer CrF$_3$ does not exhibit the $\mathcal{T}\boldsymbol{\tau}$ symmetry, the bilayer cannot exhibit this symmetry. The $\mathcal{C}_{2z}$ symmetry can be broken by sliding. To satisfy the symmetry requirement of altermagnetism, different magnetic sublattices must be connected by a symmetry operation. We consider an in-plane twofold rotation $\mathcal{C}_{2\alpha}$, with $\alpha$ taken along a generic in-plane direction, excluding the high-symmetry $x$ and $y$ axes. Clearly, $\left(\{\mathcal{T}\|\mathcal{C}_{2\alpha}\,|\,\mathbf{0}\}\right)^2 \in G_S$. We choose ${}^{c}\hat{O} = \{\mathcal{T}\|\mathcal{C}_{2\alpha}\,|\,\mathbf{0}\}\{E\|\mathcal{C}_{2y}\,|\,\mathbf{0}\} = \{\mathcal{T}\|\mathcal{C}^{\beta}_{z}\,|\,\mathbf{0}\}$, 
where $\mathcal{C}^{\beta}_{z}$ denotes a rotation by an angle $\beta$ about the $z$ axis. Fig. \ref{fig:figure2}(b) shows the crystal structure for $\beta = 11.43^\circ$, which exhibits altermagnetism. Furthermore, we find that in this bilayer system the symmetry operation connecting different magnetic sublattices is only an in-plane twofold rotation. Choosing ${}^{c}\hat{O} = \{E\|\mathcal{C}^{11.43}_{z}\,|\,\mathbf{0}\}$ can break this symmetry, leading to a fully compensated ferrimagnetic state, as illustrated in Fig. \ref{fig:figure2}(c). A more detailed discussion is provided in the Supplemental Material. First-principles calculations show that the fully compensated ferrimagnetic state is energetically more favorable than the altermagnetic state. Therefore, we further calculate the band structure of the bilayer with the stacking operation $\{E\|\mathcal{C}^{11.43}_{z}\,|\,\mathbf{0}\}$ without SOC, as shown in Fig. \ref{fig:figure2}(d), using the high-symmetry path indicated in Fig. \ref{fig:figure2}(e), and confirm a vanishing net magnetization. These results verify the fully compensated ferrimagnetism.

\textit{Summary---}In summary, we propose a general magnetic stacking theory for two-dimensional magnets with all possible magnetic configurations and systematically enumerate all 448 collinear spin layer groups. 
Using CrF$_3$ as an example, we demonstrate how magnetic stacking theory can be used to design and manipulate unconventional magnetism. 
Beyond explaining previously reported stacking-induced magnetic phenomena, our theory offers a unified and predictive guideline for discovering and engineering unconventional magnetism in 2D magnetic materials. Our work highlight stacking as a vital degree of freedom for magnetic symmetry in 2D magnets, thereby enriching the theoretical understanding of magnetism and symmetry breaking in two dimensions.

\textit{Acknowledgments---}This work is financially supported by the National Natural Science Foundation of China (Grant No. 12474229). This work is partially supported by High Performance Computing Platform of South China University of Technology.

\nocite{*}
\bibliography{ref}

\end{document}